# Lattice Mismatch-Driven In-Plane Strain Engineering for Enhanced Upper Critical Fields in Mo$_2$N Superconducting Thin Films

*Aditya Singh[1], Divya Rawat[1], Victor Hjort[2], Abhisek Mishra[3], Arnaud le Febvrier[4], Subhankar Bedanta[3], Per Eklund[2,4] \*, and Ajay Soni[1] \**

[1]School of Physical Sciences, Indian Institute of Technology Mandi, Mandi 175005, Himachal Pradesh, India.
[2]Thin Film Physics Division, Department of Physics, Chemistry, and Biology (IFM), Linköping University, Linköping, SE-58183, Sweden.
[3]Laboratory for Nanomagnetism and Magnetic Materials (LNMM), School of Physical Sciences, National Institute of Science Education and Research (NISER), An OCC of Homi Bhabha National Institute (HBNI), Jatni 752050 Odisha, India.
[4]Department of Chemistry - Ångström Laboratory; Inorganic Chemistry, Uppsala University, Uppsala, 75105, Sweden

Emails: per.eklund@kemi.uu.se and ajay@iitmandi.ac.in

Transition metal nitrides are a fascinating class of hard coating material that provide an excellent platform for investigating superconductivity and fundamental electron-phonon (*e-ph*) interactions. In this work, the structural, morphological, and superconducting properties have been studied for Mo$_2$N thin films deposited via direct current magnetron sputtering on c-plane Al$_2$O$_3$ and MgO substrates to elucidate the effect of internal strain on superconducting properties. High-resolution X-Ray diffraction and time-of-flight-elastic recoil detection analysis confirms the growth of single-phase Mo$_2$N thin films exhibiting epitaxial growth with twin-domain structure. Low-temperature electrical transport measurements reveal superconducting transitions at ~ 5.2 K and ~ 5.6 K with corresponding upper critical fields of ~ 5 T and ~ 7 T for the films deposited on Al$_2$O$_3$ and MgO, respectively. These results indicate strong type-II superconductivity and the observed differences in superconducting properties are attributed to substrate-induced strain, which leads to higher *e-ph* coupling for the film on MgO substrate. These findings highlight the tunability of superconducting properties in Mo$_2$N films through strategic substrate selection.

Superconductivity in thin films has garnered significant attention due to their well-defined geometry, enhanced surface-to-volume ratio and remarkable sensitivity to external stimuli making them ideal candidates for a diverse technological applications including single-photon detectors, superconducting quantum interference devices and quantum computing. [1, 2] The ability to tune the superconductivity in epitaxial thin films, in comparison to the bulk counterparts, enables precise control over critical parameters such as carrier density, conformality to curved surfaces thus allowing functionalities like manipulation of magnetic flux and switching in Josephson junctions. [3] Among the various material systems explored for thin film functionalities, the Transition Metal nitrides (TMNs) stand out due to their multidisciplinary array of technological applications arising from their thermal stability, refractory character and chemical resistance. [4-6] Epitaxial growth of thin film of TMNs on various substrates allows for precise control of morphology and structures at atomic-level, further enhancing their potential for advanced quantum technologies. TMNs display a range of functional properties, including superconductivity, [7, 8] quantum magnetism, [9, 10] exceptional hardness [11-13] making them suitable for applications in electronics, energy storage, photocatalysis, supercapacitor electrodes and beyond.[5, 13]

Within the TMN family, niobium, scandium and titanium nitrides have been widely studied for their superconducting behaviour, thermal managements and compatibility with device integration.[14-16] However, molybdenum nitrides (Mo–N) present a more complex case owing to their rich phase diagram, where even minor changes in temperature or nitrogen content can lead to multiple competing phases.[17] Previous research on $Mo_2N_x$ systems have explored the impact of nitrogen and argon content in the plasma during the sputtering on the superconducting properties,[18] magnetization behaviour of different $Mo_2N_x$ phases,[19] thickness-dependent superconducting properties of $Mo_2N$ films grown on AlN-buffered Si substrates,[20] surface-induced suppressed superconductivity using low-temperature tunnelling spectroscopy.[21] Despite these efforts, comprehensive investigations on phase-pure $Mo_2N$ thin films grown on substrates such as $Al_2O_3$ (MN/$Al_2O_3$) and MgO (111) (MN/MgO) is lacking, particularly where the detailed structural and low-temperature studies are involved. Moreover, recent theoretical predictions, [22] using Eliashberg formalism estimate a superconducting critical temperature ($T_c$) ~ 15.8 K for $Mo_2N$ with a strong *e-ph* coupling constant of ~ 1.2. These findings highlight the need for experimental validation, particularly since strain induced by lattice mismatch can modify the *e-ph* interactions and potentially enhance superconducting properties. In general, the in-plane compressive strain flattens the electronic bands leading to enhanced density of state near fermi energy thereby elevating the superconducting $T_c$.[23]

In this work, we report a detailed study on superconductivity in phase-pure cubic $Mo_2N$ thin films grown on c-plane $Al_2O_3$ and MgO (111) substrates, via DC reactive magnetron sputtering. The films exhibit a well-defined twin domain structure, as confirmed by X-ray diffraction (XRD) and pole figure analysis. Composition and stoichiometry are determined using Time-of-Flight Elastic Recoil Detection Analysis (ToF-ERDA), affirming near-ideal nitrogen incorporation. [the details are presented in supporting information Figure S1]

Temperature dependent electrical transport measurements reveal superconducting transitions with $T_c$ up to ~ 5.5 K. The upper critical field ($H_{c2}$) for the MN/MgO shows a significant enhancement (~ 25%) from MN/Al$_2$O$_3$. Despite having same composition, the enhancement is related to the effect of in-plane compressive strain on the MN/MgO as compared to the in-plane tensile strain on the MN/Al$_2$O$_3$, which leads to the modulation of cooper pair formation. Magnetic measurements provide the details of lower critical fields ($H_{c1}$), coherence lengths, and penetration depths, which supports the robust type -II superconducting character of the films. Additionally, the Debye temperature of the thin films is estimated from fitting of thermal response of resistance with Bloch-Grüneisen model, and suggests a critical insight into *e-ph* interactions. These results demonstrate that epitaxial MN/MgO can serve as a viable platform for superconducting applications with tunable physical properties.

**Results and Discussion:**

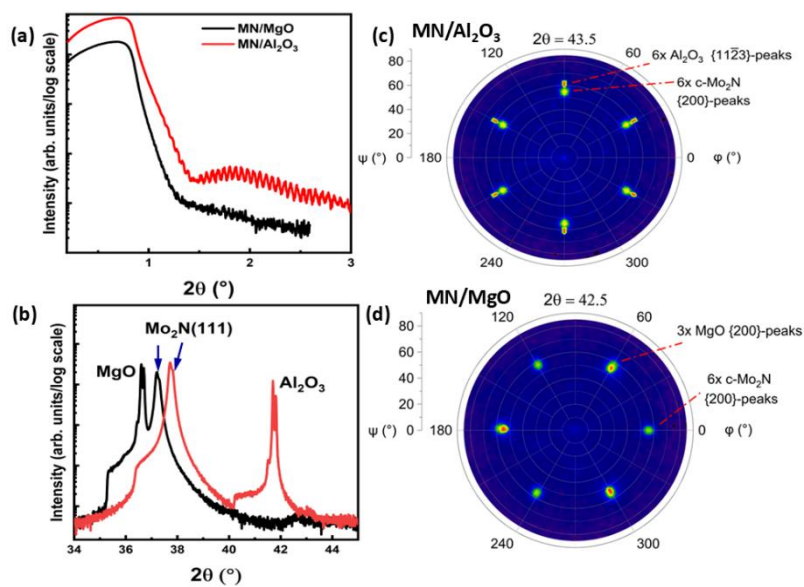

Figure 1 (a) XRR plot for MN/Al$_2$O$_3$ and MN/MgO (XRR is shifted vertically for visual clarity), (b) θ-2θ XRD patterns. Pole figures indicating single phase epitaxial growth for (c) MN/Al$_2$O$_3$ and (d) MN/MgO.

The composition of the films is determined from the ERDA (Figure S1 and Table S1 in Supporting Information file), and found to be Mo$_2$N$_{0.96}$ and Mo$_2$N$_{0.97}$ for MN/Al$_2$O$_3$ and MN/MgO, respectively, which is very close to the stoichiometric Mo$_2$N phase, within the error limits. The thickness of the films is estimated by X-Ray Reflectivity (XRR) (Figure 1(a)) and are found to be ~ 130 ± 5 nm with estimated roughness of ~ 2 nm and ~ 5 nm for MN/Al$_2$O$_3$ and MN/MgO, respectively. The mass density of the films extracted from XRR are presented in table S1 in Supporting Information. Figure 1(b) displays the XRD patterns of the two samples, where the substrate peaks for (0001) Al$_2$O$_3$ and (111) MgO are visible at 2θ ~ 41.7° (PDF 00-046-1212) and at 2θ ~ 36.6° (PDF 01-071-6452), respectively. The peak at 2θ ~ 37.7° (d-spacing 2.38 Å) for MN/Al$_2$O$_3$ and 2θ ~ 37.2° (d-spacing 2.42 Å) for MN/MgO corresponds to the (111) plane of cubic Mo$_2$N with a dominant out-of-plane orientation. The shift in the intense MN-111 peak is because of the substrate driven stress. The in-plane orientation of the films is further demonstrated by the pole figure analysis of the 200-reflection of Mo$_2$N, observed at 2θ ~ 43.5° for MN/Al$_2$O$_3$ and ~ 42.5° for MN/MgO, as shown in Figure 1(c-d). The six poles of Al$_2$O$_3$ 11$\bar{2}$3, at Ψ = 61.5° and six poles of c-Mo$_2$N 200, at Ψ = 55.5° in Figure 1(c) indicates epitaxial twin-domain growth, a common feature for cubic structures on both the substrates.[24-26] In Figure 1 (d) three prominent poles are observed at Ψ = 54.7°, corresponding to MgO 200-reflections from the substrate alongside six poles from the Mo$_2$N (200) planes, spaced every 60° in φ, with some overlapping the MgO poles. Since both MgO and Mo$_2$N are cubic, their 200 poles are expected to align at Ψ = 54.7° in a stress-free state. However, the Mo$_2$N poles appear at a slightly higher Ψ (~ 55.5°), indicating in-plane compressive strain in the film consistent with observations in Figure 1(b). The epitaxial relationship of Mo$_2$N on both substrates is as follows: (111)$_{Mo2N}$ ∥ (0001)$_{Al2O3}$ (out-of-plane) and [11$\bar{2}$]$_{Mo2N}$ ∥ [10$\bar{1}$0]$_{Al2O3}$ (in-plane); and (111)$_{Mo2N}$ ∥ (111)$_{MgO}$ (out-of-plane) and [110]$_{Mo2N}$ ∥ [110]$_{MgO}$. The morphological studies as well as the AFM images reveal a smooth surface with voids and grain boundaries, provided in the Supporting Information file (Figure S2 (a-d)).

Comprehensive characterization using ERDA, XRD and pole figure confirms that both films consist of a single-phase Mo$_2$N with identical composition. Epitaxial growth is observed on both c-plane Al$_2$O$_3$ and MgO (111) substrates, with twin domains present in each case. Despite the consistent composition, a notable shift in the XRD peak positions is observed between the two films, attributed to strain induced by their epitaxial relationship with the respective substrates. Comparing with the XRD peaks of stress-free Mo$_2$N (37.4°) as a reference, the MN/Al$_2$O$_3$ film exhibits a peak shift indicative of in-plane tensile strain, while the MN/MgO film shows a shift consistent with in-plane compressive strain.

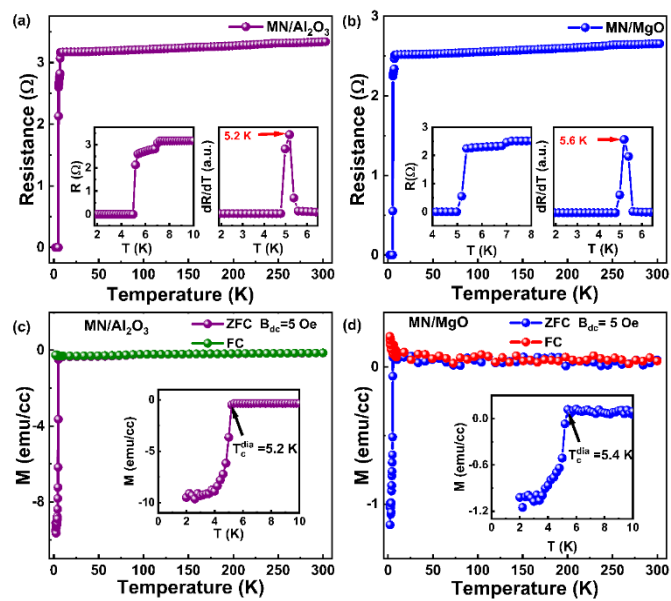

Figure 2. Temperature dependent resistance of (a) MN/Al$_2$O$_3$ and (b) MN/MgO. The inset shows d$R$/d$T$ plots for estimation of $T_c$, and temperature dependent magnetization of (c) MN/Al$_2$O$_3$ and (d) MN/MgO. The $T_c$ in insets is supporting the observation in transport measurements with in the error limits of the techniques.

The onset of superconductivity in the thin films is estimated from electrical transport and magnetization measurements. The *R* (300 K) for the films is ~ 3.33 Ω and ~ 2.65 Ω for MN/Al$_2$O$_3$ and MN/MgO, respectively, which is attributed to the increased grain boundary scattering in the MN/Al$_2$O$_3$ as compared to MN/MgO (Figure S2(c-d)). In order to study the quality of thin films, the residual resistivity ratio, $RRR = \frac{\rho\ (300\ K)}{\rho\ (6K)}$, is calculated for both the thin films which is found to be ~ 1.22 for MN/Al$_2$O$_3$ and 1.14 for MN/MgO. The significantly higher *RRR* compared to previously reported Mo$_2$N thin films on Si substrates (*RRR* ~ 0.9) [20], indicates a superior crystallinity, reduced defect density, and larger grain sizes with fewer macroscopic imperfections. The *R-T* data (Figure 2 (a-b)) confirms the metallic nature of the films with the $T_c$ estimated from the derivative plot of resistance (inset of Figure 2 (a-b)), which shows the $T_c$ ~ 5.2 K (for MN/Al$_2$O$_3$) and $T_c$ ~ 5.6 K (for MN/MgO). The estimated $T_c$ is further corroborated by magnetization measurements of the Meissner effect (Figure 2(c-d)). Magnetization data *(M-T)* obtained using both field-cooled (FC) and zero-field-cooled (ZFC) protocols exhibit robust diamagnetic behaviour with a $T_c^{dia} = 5.2\ K$ and $T_c^{dia} = 5.4\ K$, for MN/Al$_2$O$_3$ and MN/MgO, respectively. Furthermore, the *M-T* data in both ZFC and FC protocols are separated thus confirming the presence of flux pinning in the films.[27, 28] In order to rule out the possibility of substrate induced superconductivity, *M-T* measurements of bare substrates (Figure S3 (a-b)) are performed, showing no signs of superconductivity. The higher $T_c$ for the MN/MgO film as compared to MN/Al$_2$O$_3$ is associated with the compressive in-plane strain. [29] Two-step superconducting transitions, (inset of Figure 2 (a-b)), commonly arise from applied forces such as pressure, magnetic fields, strain, proximity effects and is observed in several other thin films as well. [30-32] The temperature dependence of resistivity above the $T_c$ (9 K-200 K) is analysed using the Bloch Grüneissen (BG) model, [33] by taking account of the *e-ph* interactions. The fitting of the resistivity data (Figure S4) strongly suggests that *e-ph* interactions are the dominant scattering mechanism for Mo$_2$N thin films.[34] The *e-ph* coupling strength, estimated using McMillan's relation[35] with a Coulomb pseudopotential μ$^*$=0.13, is found to be ~ 0.75 for MN/Al$_2$O$_3$ and ~ 0.76 for MN/MgO (Section D in Supporting Information). This result highlights the enhancement of *e-ph* interactions due to compressive strain in MN/MgO compared to the tensile strain in MN/Al$_2$O$_3$.

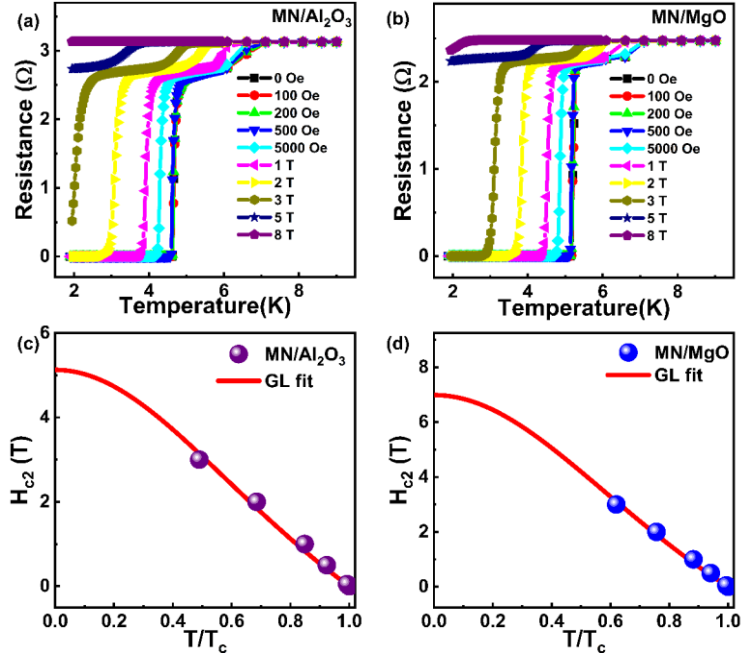

Figure 3: Low temperature magnetoresistance measurements for (a) MN/Al$_2$O$_3$ and (b) MN/MgO. Upper critical field with Ginzburg-Landau fitting of (c) MN/Al$_2$O$_3$ and (d) MN/MgO.

Low-temperature magnetoresistance measurements (Figure 3 (a-b)) revealed an increasing $T_c$ width with applied magnetic field, which is a characteristic signature of type-II superconductors.[36] The upper critical field $H_{c2}$, is estimated suing the $T_c^{onset}$ criterion, with the normal state resistance taken at ~ 6.1 K and by fitting the data with the Ginzburg- Landau (GL) model with an extrapolation to determine the $H_{c2}(0)$ for both films. According to the GL model, the critical field at a temperature, $T$ is given by $H_{c2}(T) = H_{c2}(0)\frac{(1-t^2)}{(1+t^2)}$, where $t$ is the reduced temperature given by $t = \frac{T}{T_c}$ and $H_{c2}(0)$ is critical field at 0 K. Based on this model, the estimated $H_{c2}(0)$ is found to be ~ 5.1 T for MN/Al$_2$O$_3$ and ~ 7 T for MN/MgO, which is more than 25% enhancement in the MN/MgO then MN/Al$_2$O$_3$.

The enhancements is attributed to the increased flux pinning potential resulting from a compressive in-plane strain on MN/MgO in contrasts with the tensile in-plane strain in MN/Al$_2$O$_3$ and also correlates with enhanced the *e-ph* interactions.[23, 37, 38] To understand further on the nature of the superconductivity and the limiting mechanism of the copper pair formation, it is important to dive deeper into the mechanism governing $H_{c2}$. For this, the Werthamer-Helfand-Hohenberg (WHH) formula $H_{c2}^{orb}(0) = -0.693 T_c \left(\frac{dH_{c2}}{dT}\right)_{T=T_c}$, [39] is used for estimation of the zero-temperature orbital limited upper critical field, $H_{c2}^{orb}(0)$. The calculated values are ~ 4.4 T for MN/Al$_2$O$_3$ and ~ 5.8 T for MN/MgO. These estimates are lower than the $H_{c2}$ estimated from GL fitting. In general, the upper limit on $H_{c2}$ is given by the Pauli Paramagnetic limit, which is given by $H_p = 1.86 T_c$ (9.6 T for MN/Al$_2$O$_3$ and 10.3 T for MN/MgO). The estimation and analysis clearly indicates a conventional BCS type superconductivity.[40]. To understand the mechanisms of Cooper pair breaking in type-II superconductors under an applied magnetic field, both orbital and spin-paramagnetic effects on the $H_{c2}$ are considered. According to the Maki theory, the Maki parameter ($\alpha$), which quantifies the relative strength of these effects, is defined as $\alpha = \sqrt{2} \frac{H_{c2}^{orb}(0)}{H_p(0)}$,[41]. The $\alpha$ values are found to be ~ 0.64 (for MN/Al$_2$O$_3$) and ~ 0.8 (MN/MgO), both less than 1, confirming that the superconductivity in Mo$_2$N thin films is limited by orbital depairing mechanisms. The GL coherence length ($\xi_{GL}$), is calculated by $\xi_{GL} = \sqrt{\frac{\phi_0}{2\pi H_{c2}(0)}}$, where $\phi_0 = 2.068 \times 10^{-15}$ T.$m^2$ is magnetic flux quantum. The coherence length is found out to be ~ 8.01 nm and ~ 6.85 nm for MN/Al$_2$O$_3$ and MN/MgO films, respectively.

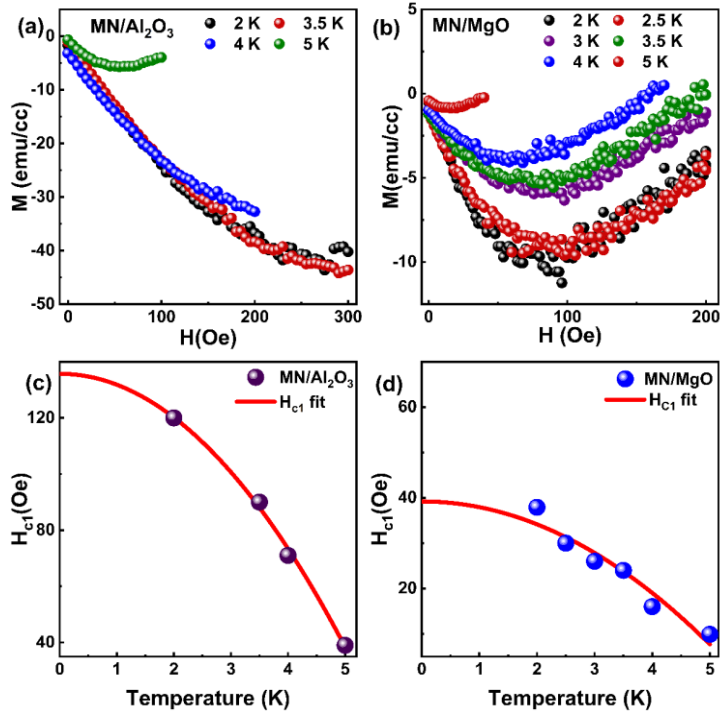

Figure 4: *M-H* plots (a-b) and $H_{c1}$ vs T (c-d) plots of MN/Al$_2$O$_3$ and MN/MgO respectively.

Table 1: Superconducting parameters of MN/Al$_2$O$_3$ and MN/MgO.

| Sample | Thickness (nm) | $T_c$ (K) | $H_{c1}$ (Oe) | $H_{c2}(0)$ (T) | $H_{c2}^{orb}(0)$ (T) | $H_c$ (Oe) | $\xi_{GL}$ (nm) | $\lambda$ (nm) | $\alpha$ |
|---|---|---|---|---|---|---|---|---|---|
| MN/Al$_2$O$_3$ | 130 – 135 | 5.18 | 135 | 5.12 | 4.37 | 1469 | 8.01 | 197 | 0.64 |
| MN/MgO | 130 – 135 | 5.57 | 39 | 6.98 | 5.83 | 814 | 6.85 | 416 | 0.8 |

In order to estimate the $H_{c1}$ of the samples we have used the Meissner–Ochsenfeld criterion on the *M-H* measurements below $T_c$ (Figure 4 a-b).[42] The obtained $H_{c1}$ at different temperatures is analysed using the equation $H_{c1} = H_{c1}(0)[1 - \left(\frac{T}{T_c}\right)^2]$, where $H_{c1}(0)$ is lower critical field at 0 K. The obtained $H_{c1}(0)$ values for MN/Al$_2$O$_3$ and MN/MgO are found to be ~ 135 Oe and ~ 39 Oe, respectively. Further we have estimated the superconducting penetration depth ($\lambda_{GL}$) by $H_{c1} = \frac{\phi_0}{4\pi\lambda_{GL}^2}\ln\left(\frac{\lambda_{GL}}{\xi_{GL}}\right)$, and $\lambda_{GL}$ is found to be ~ 197 nm and ~ 416 nm for MN/Al$_2$O$_3$ and MN/MgO respectively. The GL parameter ($\kappa_{GL}$) given by $\kappa_{GL} = \frac{\lambda_{GL}}{\xi_{GL}}$ is found to be ~ 24.6 and ~ 60.7 for MN/Al$_2$O$_3$ and MN/MgO, respectively. The values are much greater than $\frac{1}{\sqrt{2}}$, indicating a strong type-II superconductivity. To assess the superconducting condensation energy, the thermodynamical critical field ($H_c$), is evaluated using the relation $H_{c1}(0)H_{c2}(0) = H_c^2 \ln\kappa_{GL}$ and found to be ~ 1469 Oe for MN/Al$_2$O$_3$ and ~ 814 Oe for MN/MgO. All the characteristic superconducting parameters for both thin films are tabulated in Table 1.

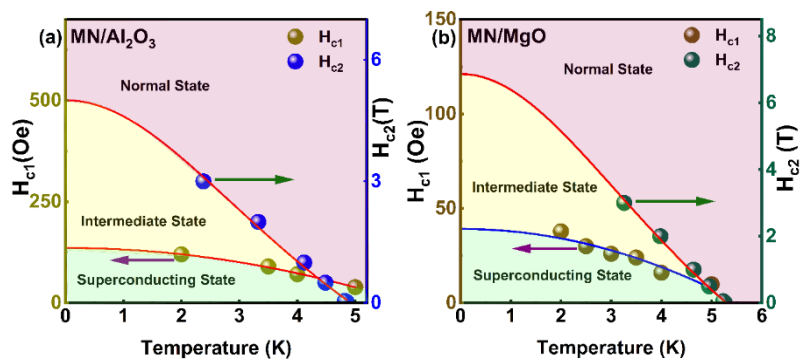

Figure 5. Phase diagram showing the superconducting, intermediate, and normal state of the (a) MN/Al$_2$O$_3$, (b)MN/MgO.

Considering both $H_{c1}$ *(T)* and $H_{c2}$ *(T)* for the films, the superconducting phase diagram is presented in Figures 5(a-b), showing distinct superconducting, intermediate, and normal states. The detailed analysis of the phase diagram confirms the thin films to be a strong type-II superconductors. These findings highlight the potential of $Mo_2N$ thin films for integration into next-generation superconducting quantum and power electronic devices.

**Conclusion:**

In summary a significant enhancement of ~ 25% in the $H_{c2}$ of MN/MgO is attributed to *e-ph* modulation arising from compressive in-plane strain, which enhances flux-pinning potential and overall superconducting performance compared to the tensile-strained MN/$Al_2O_3$. These findings demonstrate the crucial role and impact of strain engineering in tuning the superconducting behavior of transition metal nitride thin films. This work pave the way for their applicaiton in kinetic inductance devices, superconducting qubits and miniaturization of superconducting interconnects for next-generation electronic technologies, highlighting the significant impact of strain on the superconducting properties of $Mo_2N$ films.

**Experimental Details:**

Thin film deposition: Molybdenum nitride thin films were grown, on sapphire ($Al_2O_3$ (0001)) and MgO (111) substrates, by DC reactive magnetron sputtering in an ultrahigh vacuum chamber.[43] Before deposition, the MgO substrates were cleaned by multiple sonication cycles first in (i) a Hellmanex soap solution (~ 3 mins), followed by (ii) deionized water (two times for ~ 5 mins), (iii) acetone ( ~ 10 mins), (iv) ethanol (~ 10 mins) and finally dried with compressed nitrogen gas.[44] This procedure helps in the removal of possible hydroxides and the carbonates present on the surface. The substrates were kept in the deposition chamber and evacuated to a pressure of ~ $8.9 \times 10^{-9}$ Torr and the substrate temperature was kept at ~ 500° C. The deposition of $Mo_2N$ was carried out with a mixture of Argon and Nitrogen (flow rates of Ar ~22 sccm and $N_2$ ~ 33 sccm).

Characterisation Studies: The compositions of the films were determined by ToF-ERDA using ~ 36 MeV $^{127}I^{9+}$ ions as probing beam with a incidence angle ~ 67.5°, and a recoil angle ~ 45°. [45] The data were analysed using the Potku code.[46] The crystal structure was investigated by XRD using a PANalytical X'Pert Pro diffractometer in Bragg-Brentano geometry and equipped with a Cu $K_α$ source operated at ~ 45 kV and ~ 40 mA. The incident optics had a ~ 0.5° divergence slit and a 0.5° anti-scatter slit, and the diffracted optics included a ~ 5.0 mm anti-scatter slit, a 0.04 rad Soller slit, a Ni-filter, an X'Celerator detector measuring in ~ 0.008° steps sizes with an equivalent counting time of ~ 19 s per step. For XRR, the PANalytical X'Pert Pro diffractometer was operated in line-mode with a hybrid mirror module with ~ 0.5° divergence slit for incident optics and a ~ 0.125° divergence slit for the diffracted optics. For pole figure, the instrument was operated in point-mode with crossed-slit module set at 2 x 2 mm as incident optics, and a ~ 0.27° parallel plate collimator for diffracted optics. Scanning electron microscopy (SEM) was performed using a Zeiss Sigma 300 with the field emission gun was operated at 2 kV. The atomic force microscopy (AFM) images of the films were taken using Bruker Dimension icon AFM.

Physical and Magnetic properties measurements: The low temperature electrical transport properties were performed using the physical property measurement system (PPMS, Quantum Design make). The temperature dependent magnetization measurements were carried out using using an MPMS3 SQUID-VSM magnetometer (Quantum Design make). In the zero-field-cooled (ZFC) protocol, the sample was first cooled from room temperature down to ~ 2 K in the absence of external magnetic field and the magnetization was recorded during the subsequent warming cycle in the presence of a ~ 5Oe field. For the FC measurement, the sample was cooled in the presence of the same constant external magnetic field and the magnetization was recorded continuously during this cooling process.

**Acknowledgement:** AS acknowledge IIT Mandi for research facilities and DST India for Indo-Sweden bilateral grant (Grant No. DST/INT/SWD/VR/P-18/2019). PE acknowledges funding from the Swedish Government Strategic Research Area in Materials Science on Functional Materials at Linköping University (Faculty Grant SFO-Mat-LiU No. 2009 00971), the Knut and Alice Wallenberg foundation through the Wallenberg Academy Fellows program (KAW-2020.0196, P.E.), and the Swedish Research Council (VR) under Project No. 2021-03826. AM and SB thank the Department of Atomic Energy (DAE), Government of India for the financial support.

**Data Availability**

The data that support the findings of this study are available from the corresponding author upon reasonable request.

<h1 style="text-align:center;">Supporting Information</h1>

# Lattice Mismatch-Driven In-Plane Strain Engineering for Enhanced Upper Critical Fields in Mo$_2$N Superconducting Thin Films

*Aditya Singh[1], Divya Rawat[1], Victor Hjort[2], Abhisek Mishra[3], Arnaud le Febvrier[4], Subhankar Bedanta[3], Per Eklund[2,4] *, and Ajay Soni[1] **

[1]School of Physical Sciences, Indian Institute of Technology Mandi, Mandi 175005, Himachal Pradesh, India.
[2]Thin Film Physics Division, Department of Physics, Chemistry, and Biology (IFM), Linköping University, Linköping, SE-58183, Sweden.
[3]Laboratory for Nanomagnetism and Magnetic Materials (LNMM), School of Physical Sciences, National Institute of Science Education and Research (NISER), An OCC of Homi Bhabha National Institute (HBNI), Jatni 752050 Odisha, India.
[4]Department of Chemistry - Ångström Laboratory; Inorganic Chemistry, Uppsala University, Uppsala, 75105, Sweden
Emails: per.eklund@kemi.uu.se and ajay@iitmandi.ac.in

The supporting information has the additional details of characterization and experiments complementing to the main text.

### (A) Time-of-Flight Elastic Recoil Detection Analysis (ToF-ERDA):

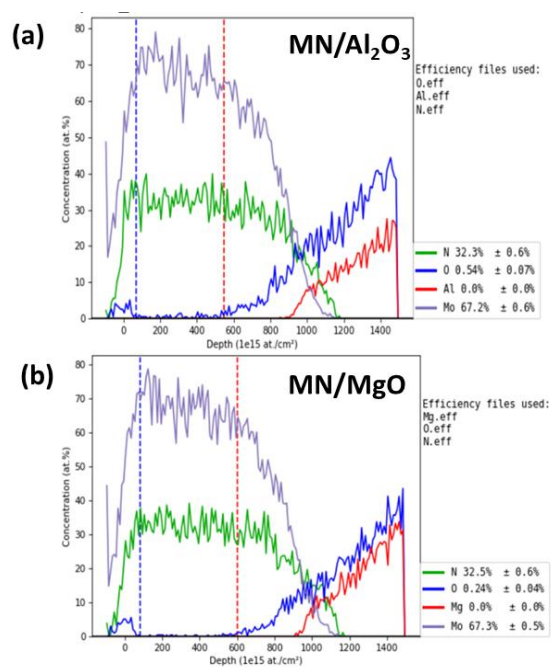

Figure S1: ToF-ERDA of (a) MN/Al$_2$O$_3$ and (b) MN/MgO.

The ERDA depth profiles (Figure S1. (a-b)) for MN/$Al_2O_3$ and MN/MgO show uniform elemental distributions of N and Mo across the measured depth range. No significant contribution of Al or Mg is detected in the measured depth range, which can be expected to be coming from the substrates. In the fitting model for XRR, a very thin layer of $3.4 \pm 1$ nm of oxide is used for the convergence of fitting parameter and is considered to be present on both films due to post-oxidation, once the sample is out of the deposition chamber. The elemental composition of the films along with the density and roughness as extracted from the ERDA and XRR measurements are detailed in Table S1.

Table S1. Elemental composition, density, and roughness of the films.

| Films/Substrate | Composition from ERDA, ($\pm 3\%$) | | Density and Roughness from XRR | |
|---|---|---|---|---|
| | $Mo_2N_x$ | O-contamination | Density [g cm$^{-3}$] | Roughness (nm) |
| MN/$Al_2O_3$ | $Mo_2N_{0.96}$ | 0.6 % | $8.9 \pm 0.1$ | $2 \pm 1$ |
| MN/MgO | $Mo_2N_{0.97}$ | 0.2 % | $8.8 \pm 0.1$ | $5 \pm 2$ |

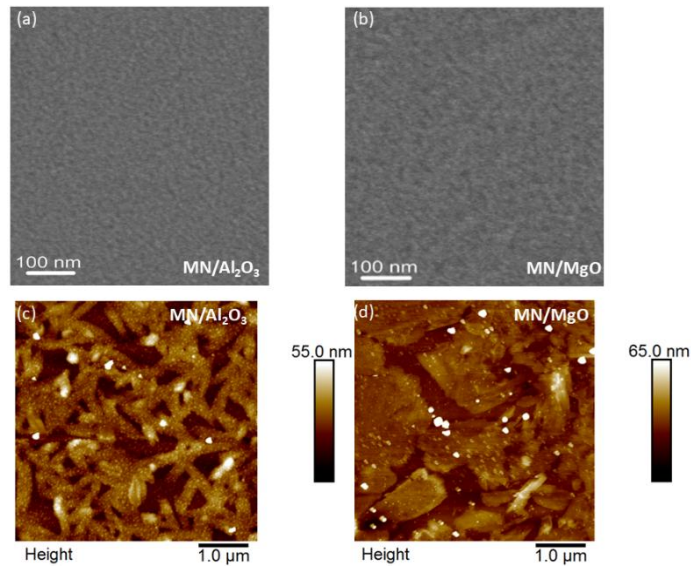

Figure S2. SEM images of (a) MN/$Al_2O_3$ and (b) MN/MgO. AFM images of (c) MN/$Al_2O_3$ and (d) MN/MgO.

### (B) SEM and AFM analysis:

Figure S2 (a-b) shows SEM images of films, where MN/$Al_2O_3$ has a homogeneous arrangement of grains with less volume of grain-boundaries while MN/MgO has larger grains with ridges and voids reflecting higher roughness and lower density, which is also in line with the estimation from the XRR. The observations of higher grain boundaries in $Al_2O_3$ are also supported by the AFM images in Figure S2 (c-d).

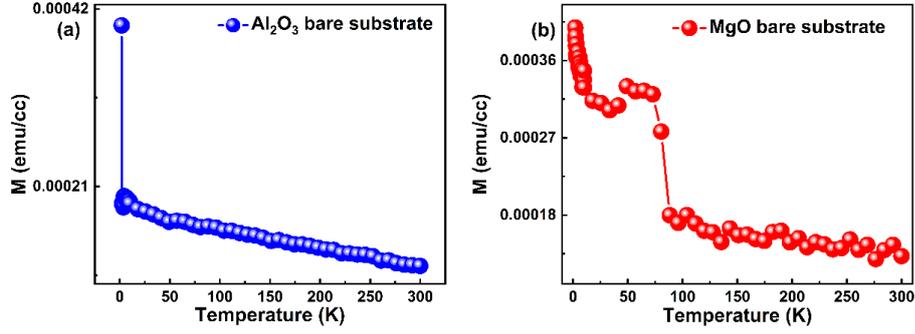

Figure S3: M-T of bare substrates (a) $Al_2O_3$, (b) MgO.

### (C) Magnetisation vs Temperature plots of the bare substrates.

The M–T plots (Figure S3 (a–b)) of the bare substrates exhibit no signatures of superconductivity, thereby confirming that the observed superconducting behaviour is arising from the films and not from the substrates.

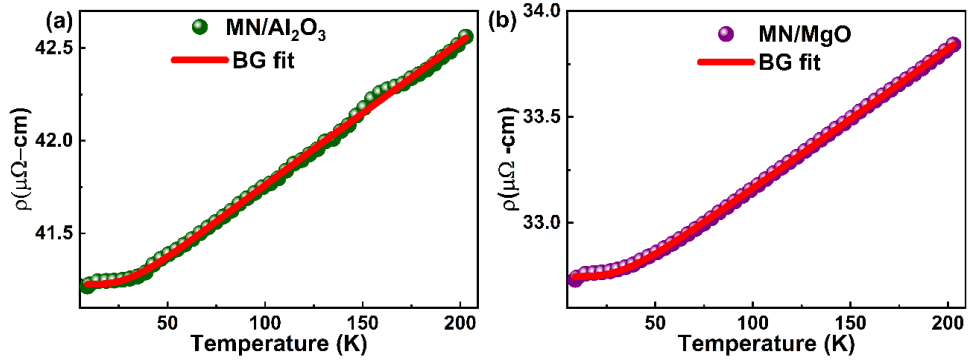

Figure S4: Bloch-Grüneisen fitting of resistivity for (a) MN/$Al_2O_3$ and (b) MN/MgO.

### (D) Resistivity data analysis using Bloch-Grüneisen equation and electron phonon coupling strength estimation using McMillan's relation

The temperature dependent resistivity ($\rho(T)$) (Figure S4 (a-b)) have been fitted with the Bloch-Grüneisen (BG) model, given by $\rho(T) = \rho_0 + \alpha_{el-ph}(\frac{T}{\Theta_R})^n \int_0^{\frac{\Theta_R}{T}} \frac{x^n}{(e^x-1)(1-e^{-x})} dx$, where $\rho_0$ corresponds to residual resistivity, $\Theta_R$ is Debye temperature, whereas $\alpha_{el-ph}$, and $n$ are material specific constants. The fitting revealed $\Theta_R \sim 194$ K for MN/$Al_2O_3$ and $\Theta_R \sim 201$ K for MN/MgO, while the exponent $n \approx 5$, which is a signature of cleaner electronic path like in metals.[1, 2] The electron-phonon coupling strength ($\lambda_{el-ph}$) is calculated using the McMillan formula given by $\lambda_{el-ph} = \frac{-1.04 + \mu^* \ln(1.45\, T_c/\Theta_R)}{((1-0.62\mu^*)\ln(1.45\, T_c/\Theta_R)) + 1.04}$ where $\mu^*$ is the repulsive screened Coulomb pseudopotential which is taken as 0.13, as in the case of transition metal based compounds.[3] Using the $T_c$ and $\Theta_R$ the values of $\lambda_{el-ph}$ have been estimated to be ~ 0.75 (for Mn/Al$_2$O$_3$) and ~ 0.76 (for MN/MgO). These results support the observations of in-plane compressive strain in Mn/MgO enhances the electron-phonon interactions compared to the tensile strain in Mn/Al$_2$O$_3$.